# From Agile to DevOps, Holistic Approach for Faster and Efficient Software Product Release Management

Fatih Bildiri[a*], Ömür Akdemir[a]

[a]Business School, Management Department, Ankara Yildirim Beyazit University, Ankara, Turkey.

## Abstract

Release management is one of the most important software processes and is a set of processes that includes the compilation, configuration, and management of software versions in different environments. In recent years, changes in processes, technologies, and tools and changes in practices and understanding have paved the way for more effective, efficient, sustainable, reusable models and methods in this field. The purpose of this study is to examine the DevOps idea to produce a flow, highlight their benefits, and investigate with a model how these philosophies, which are two of the most important processes and methods in software development today, can reveal an effective release management process. What has been learned from the research is how the agile and DevOps practices, which have become widespread in recent years, can be positioned in a general flow in the release management process, although there are different practices, flows, disciplines, and technology. Sharing a case study on these issues in future studies and an experience sharing research where the flow is applied as a case study will reveal positive feedback on the real-life application and results of the flow and the model. Further, a literature review studies in which deficiencies in the literature are identified will be useful in determining the gaps in the process.

## 1. INTRODUCTION

Software processes are both defined by many international standards and are widely used today with their practices. Release Management, which is included in all of these international standards (such as ISO 27001, 27005, 20000, 31000, ITIL V3, CMMI, etc.) in software processes and includes the practices of publishing the code, product, software after the development phase, is a very critical phase. The basic principles underlying this challenge are complicated by differences in business processes, differences in environments and services, and even differences in standards. Also, especially developing technologies, emerging new approaches, new services, and the increase of companies that develop solutions in this field make these processes very open to change and development, which creates the need for frequent updating of models. The first of these, which emerged with a manifesto written in 2001 is one of the most important methods in the software world today, Agile Manifesto (Highsmith & Fowler, 2001). The other is the DevOps idea, which

---

was put forward at the Agile Toronto conference in 2008. In this study, the DevOps idea will be examined to produce a flow, highlight their benefits, and investigate with a model how these philosophies, which are two of the most important processes and methods in software development today, can reveal an effective release management process.

*General Situation*

According to the definition of Newman (2015), the state of a system in a specific configuration is called a version. Release management is the name given to the flow that covers all of the processes in which these specific versions are defined and kept in a recordable state. However, Gordon and Hernandez (2016) define the concept of release management as a computer science discipline that is used to control the release of applications, software, updates, and patches in different domains and environments. The importance of software release management has always been high in the field of software development where versioning is very important and documentation, configuration, and even implementation are based on version. Many companies have release manager and release coordinator roles. Because actually, release management refers to the stage and this process in which the pure source code turns into a software product.

It is difficult to separate the issue of release management with agile, which is one of the most used methodologies in software development today. Because the goal of Agile development is to be able to create stable releases more often than traditional development methodologies allow. In other words, new features and functions are released at short intervals or even several versions per month instead of being developed for a long time and released in a single version, which makes the release management more critical. (Shore, 2007)

## 2. LITERATURE REVIEW

Although issues such as Agile and DevOps are newly emerging concepts, while Agile was introduced in 2001, DevOps started to come to the fore in 2008, there are studies for different disciplines in this field (Fowler & Highsmith, 2001). For example, in agile software development keyword search, approximately five thousand articles are listed in Web of Science. For DevOps, this figure approaches a hundred. However, these models and technologies are examined from the perspective of the release management, which is one of the most important steps and objectives of the software management and software development life cycle, and there is no flow suggestion that all these models can work end-to-end. This study focuses on this deficiency in the literature and aims to offer a holistic flow and model proposal.

**Software Development Life Cycle**

According to the definition of Chapple and Seidl (2020), the Software Development Life Cycle (SDLC) describes the steps in a model for software development throughout its life. In other words, every development process, especially the product development process, includes certain stages. It should not be forgotten that software is also a product. In this context, it is possible to talk about various stages for the development of a production item, a vehicle, or a product, and there are similar stages for the software. The sum of these stages is the software life cycle and the development process is called the software development life cycle. There are different approaches to this model. The first approach reduces the SDLC steps to 4 and these can be counted as follows (Blanchard and Fabrycky, 2006). These stages can be listed as planning, analysis, design, and finally adaptation and operation.

Another approach in this area is the 7-step approach presented by Kendall and Kendall. This approach consists of the following steps (Kendall and Kendall, 1992): 1. Determining the problem, opportunities, and goals 2. Determining the human level information needs 3. Determining the system needs (demands) 4. Designing a system that responds to the requests obtained in the light of the steps so far 5. Development and documentation of the software. 6. System tests and determination of maintenance steps. 7. System realization and evaluation. It is also possible to get rid of the software development life cycle from a step-based approach, and a view in which different methods come to the fore and are shaped by different principles and approaches. There are different software development life cycle models, such as the waterfall, incremental, spiral, and rapid prototyping (Pressman, 2007). In this life cycle, which can be defined as the journey of the software product from scratch to infinity, software release management also takes an important place. Besides, methods such as Agile and DevOps have been included in different stages of this cycle over time.

**Release Management and Change Management**

Release management is defined as "*a software engineering discipline that controls the release of applications, updates, and patches to the production environment. The goal of release management is to provide assurance that only tested and approved application code is promoted to production or distributed for use*" (Gordon & Hernandez, 2016). On the other hand, the coordination of the deployment of the change is the responsibility of the change management. Change management is defined as "*a set of processes that deals with the management, uploading, tracking, monitoring, and retrospective recording of all changes. It is closely related to release management and deployment processes and matches all business processes*" (Gallacher & Morris, 2012).

**Latest Methods and Principles**

The release management process has a structure that contributes greatly to the increase of service quality and continuity in IT, to decrease the risks, costs, time of product release, and to increase the consistency of the live environment. Many different technologies and tools continue to emerge to increase this contribution and efficiency. Especially release automation systems and process automation systems contribute to increasing the efficiency of release management. However, some important principles and technologies emerge throughout the entire life cycle before the release management process. These are, CI / CD, Cloud DevOps Systems, Build Automation Platforms and Processes, Agile, Pipeline Management Systems and Technologies, Deployment and Release Technologies, Snapshot and Version Management Softwares, etc.

These new methods and principles enable the release management to be handled more effectively in many areas. Keeping the configuration in version management systems such as git enables monitoring and control mechanisms. It also provides a structure that allows changes to be implemented, moved, and rolled back. Also, it provides the opportunity to manage without the need for patches in live environments, working with branches and managing different environments, real-time data management, and determining the needs and conditions of the environments can be listed as other advantages. Besides, backed up servers, real-time server management, virtual and cloud environment management also provide a very clear management opportunity. Also, pipeline logic and philosophies such as pipeline as a code provide very advantageous management. This structure means the implementation of pipelines and steps in this pipeline such as test and deploys as software components in continuous management environments. In addition to these methods, infrastructure as a code and configuration as code structures also ensure effective handling of DevOps principles and release management.

**Agile in Release Management**

Agile Software Development Method is a frequently used software development methodology. The main feature of this software development methodology is that it is easy to integrate and adapt quickly to customer notifications. One of the most valuable features of agile projects is the release of functions and the frequent release of new versions, which significantly increases the value and makes the release management process important to agile (Shore, 2007).

Overlooking the release processes while dealing with agile process implementation and agile transformation will result in huge losses. Within the Agile structure, it is a quality release management of the product that customers and stakeholders expect from the software development team. The new features of the product may be developed quickly, but it would not be correct to say that agile processes can be advanced if these innovations take days or even weeks to be brought to life and loaded into different environments. The way to get frequent and fast releases in integration with Agile is also through a proper DevOps structure. The problems in the quality control/test engineering part are also valid here. Insufficient resources make the automated start-up system imagine.

**DevOps in Release Management**

Release Management provides content packaged in a full version to the production environment and indeed different working environments. Also, the planning of the releases systematizes the management of large-scale releases and versions that we can call release candidates (Agarwal et al., 2010). Especially the management of such multimedia broadcasts, versioning, and deployment processes can become very complex processes, and the need to use different pipelines arises (Sharma, 2017). If the process starts, one of the things that is needed the most in such a process is to establish an organization and technology network that can act in an integrated manner throughout the entire life cycle of an application, from development and testing to deployment and operations. It is named DevOps. In other words, this philosophy, which follows the planning, monitoring, and development flow from the beginning to the end, by establishing a structure that enables development and operation teams to work together, and sometimes even bringing these two teams into one team, takes its name from these teams. Another important feature of DevOps is that it automates and uses applications that automate manually performed processes that are inefficient in terms of time management by incorporating different technologies and principles into the process. These automated processes and tools also enable developers and operations teams to communicate with each other, while ensuring efficiency by ensuring that interactive work is done independently.

According to Hüttermann (2012), DevOps is the job of creating a flow by closing the gap between development and operation teams throughout the entire software development process and managing the process without loss. This job reduces batch size and cycle size rates in software development projects, and also facilitates automated publication management and handling deployment and release processes together.

While DevOps is a set of philosophies and operations, many reasons make DevOps have such a positive impact on release management. These basically can be listed as Continuous Integration, Continuous Deployment, Infrastructure as a Code, Monitoring, Web Services, Log and Data Management and Institutional Structure, Communication, and Cooperation culture. One of the main factors that make the release management process successful in the DevOps perspective is to understand the dynamics and structures that make up it. For this, it will be useful to examine these structures one by one.

Continuous Integration: It is a pipeline system that enables the developed versions and codes to be integrated with the previous version by going through certain stages and tests in a pipeline logic. Continuous Deployment: Continuous deployment is the ability to deploy the desired version by manually triggering it to the desired environment. The basis is that when releasing the software, establishing a trust mechanism and developers have to trigger manually (Arundel and Domingus, 2019). Hering (2018) argued that the definition of infrastructure as a code can be expressed as an architecture in which all infrastructures connected to operations are kept in a specific configuration unit and the configuration setup is defined in a structure that can be processed sequentially. According to Janca (2020), microservice is an acronym that characterizes microservice architecture, and when we see this definition, we need to think of an architecture based on multiple services running on the same unit and used through the application programming interface. Also, corporate cooperation and culture feed the DevOps principles in terms of business conduct and processes, and tracking and retrospective records in terms of incident and log management. And all these possibilities enable the DevOps philosophy to provide a functional framework for proper release management.

**Model for Faster and Efficient Release Management**

Changes in software development processes and technology, the emergence of new applications, tools, architectures, and philosophies also reveal the need to model and define processes within a practice. When the principles are examined from the perspective of release management, it is seen that end-to-end management, process, and model are needed. Agile methods, practices, end-to-end application habits, and especially with the effect of DevOps practices, philosophy, and dynamics such as CI / CD, which have started to increase in importance and value in recent years, it is necessary to define a model that includes all of these principles and handles the software development process from end to end.

To address such a model, it is necessary to consider the entire software development lifecycle from end to end. Because in Winkler (2011) argued that the goals of version management do not only cover the release phase. These goals include effective management of all stages, from planning and calculating the relevant version, to developing the processes to be used in the project presentation, and to managing customer expectations during and after project delivery. Such a model should follow the structure below. A model with this structure can find a holistic place in the literature in terms of both implementation of quality functions, ease of configuration management, agile adaptation, and integration to DevOps processes, and can be followed in real-life scenarios with its compatibility with the flow of SDLC processes and steps.

1. Requirements and Analyzing Step: This step should include three steps. Gathering software requirements, their analysis and planning based on epics, user stories, and issues following agile principles At this stage, it is determined which of the relevant methods of the agile will be used, the collected requirements are divided into business and functional requirements, business requirements are parsed to functional requirements and epic-based plans are made on board basis.

2. Change Management: This step is the stage in which issues related to the relevant requirements are created as change requests and their approval processes are advanced by the relevant product, project manager, or product owner. Items that will enter the release process are created and approved at this stage.

3. Development and Testing: This stage is the stage where approved change requests are coded, developed, and tested. This stage is more critical in one aspect because, according to Gordon and Hernandez (2016), the release management process starts in QA testing activities. In other words, it can be stated that release management processes and criteria come into play from this stage. It is important to take into account the agile practices applied in the previous stages in the development and testing process for the healthy walking of this stage.

4. Configuration and Deployment: The most important part of the process in terms of release management is here. The agile practices, CI pipelines, CD practices and DevOps applications applied throughout the entire process are actually for the versioning of the code that has passed certain tests, verified, authorized by QA under the correct configuration, and emerged as the output of the release management process. Planning the configurations, which is the first stage of this process, is very critical. Hohmann (2003), emphasizes that the configurations that are used in the release management process, known as institutional and widely used, make the release management more effective. Sharma (2017) examines release and deployment together. It even expresses that release and deployment are referred to as the whole set of processes related to the media-based deployment of products and services that emerge as a result of an iterative process and deal with their distribution in different ways. From this perspective, the Configuration, QA Authorization, and Deployment processes stand in the middle of release management and affect it end-to-end, even the deployment is the stage where this process emerges as a product. For this reason, the practices and technologies to be applied here are important in terms of applying the release process as a model.

5. Review and Closure: It is the step that includes the monitoring and review phase and the closure of the release after the implementation of the product that emerged during the release management process in different environments. It is the last step of the release management process, in this process, steps such as evaluation, retrospective from agile practices are applied, and the implementation of the existing monitoring and plan stages within the DevOps process will make the last step of release management more healthy and functional. Besides, this stage also reveals the success of the release, so at this stage, it may be necessary to check whether the processes are fully functioning, the use of official standards and procedures, the functionality of control and acceptance gateways, and a closing audit. After explaining the model with principles and stages, we can list it as follows;

1- System requirements are collected from users and customers. These requirements are kept integrated with requirement management systems.

2- Business and software analysis processes are improved, business requirements evolve into functional requirements.

3- Sprint plans are started about the requirements, issues are designed in story and epic-sized backlog.

4- CRs i.e. change requests are created for relevant updates and defects.

5- It is determined that the CRs regarding the project manager, product manager, scrum master meetings will be approved and developed.

6- The system is coded and developed. The coding of the system is completed at this stage. At this stage, working with agile methods and taking advantage of the flexibility and iteration opportunities provided by the agile provide a serious advantage for the model.

7- Developed codes are tested. Using agile methods and DevOps practices during the test phase helps the stages such as continuous integration, development, test, deployment to progress effectively.

8- Authorization of QA, Configuration, and other stakeholders is provided.

9- The code is deployed to the relevant environment. Integration of CI/CD and DevOps processes is important at these stages.

10- Post-implementation reviews are made, certain audits are made.

11- Relevant publications and CRs are closed, the version is completed. Since this version is integrated with DevOps processes, it is very open to testability, traceability, and progress in an automated pipeline. However, it is important that these approaches and methods advance in an iterative and repetitive manner at every stage and the integration of both Agile applications and DevOps tools and technologies into processes.

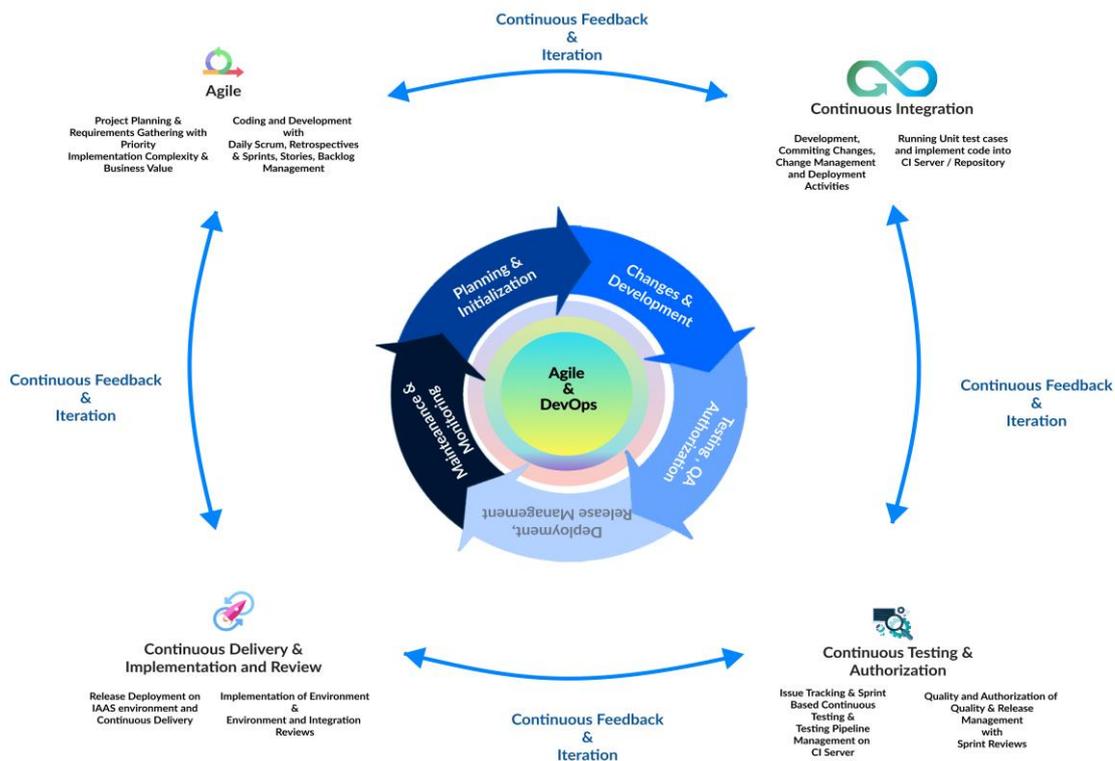

**Figure 1:** Model Suggestion with Agile and DevOps in Release Management

### 3. Advantages and Disadvantages of the Model

The most important feature of this model is that it handles change and release management processes together and while doing this, it progresses in a structure suitable for DevOps and Agile practices. After planning the flow in general terms, agile principles can be applied especially throughout the entire flow and provides a structure suitable for DevOps processes. The implementation of these practices reduces the risks in software release processes, minimizes the possibility of disruption, facilitates communication and documentation management, provides an iterative structure, providing modularity, reusability and applicability. Besides, automatic release, which is a stage of the effective implementation of DevOps practices, makes the release management process much faster, time-saving, and lean (Hüttermann, 2012).

Also, according to the results of their research and suggested in Logue and McDaid (2008), incremental and renewable release methods that can be integrated with agile methods and can be planned following the changes that occur as the project progresses contribute to efficiency. But the study by Elberzhager et al. (2017) shows that; In addition to the case study where we have seen prosperous applications of companies such as IBM and Netflix, in the case of Fujitsu, the goals, processes, questions and end-to-end pipeline must be well defined for the success of DevOps applications and the release processes managed by DevOps. Dyck et al. (2015) measure the following clearly in their study; An advanced and competent release management and engineering process is in line with the DevOps mentality and takes a highly supportive approach to it. Besides, applying the release processes within the companies in a correct management model and compatible and integrated with DevOps increases the efficiency in the production processes and thus the efficiency and value of the product. However, teams that are not competent in agile may experience serious problems with this integration. It is proclaimed in this study that providing a more flexible, more adaptable, and more fragmented structure in our study and in the model we present can facilitate this transition. Furthermore, it can be said that these processes require very serious measurement and analysis, otherwise their impact will decrease.

## 4. Acquisitions and Future Researches

This study, was mainly followed by a literature review and documenting a flow obtained from practical experience, literature, and established standards and proposing a model in which different standards and practices would be interconnected. What has been learned from the research is how the agile and DevOps practices, which have become widespread in recent years, can be positioned in a general flow in the release management process, although there are different practices, flows, disciplines, and technology. Sharing a case study on these issues in future studies and an experience sharing research where the flow is applied as a case study will reveal positive feedback on the real-life application and results of the flow and the model. Further, a literature review study in which deficiencies in the literature are identified will be useful in determining the gaps in the process.

Furthermore, it will be useful to examine different technologies and practices in this field. New management techniques that have started to emerge and their adaptations to software management have revealed new practices in software management. It will be useful to examine these practices in the context of change and release management. There is a serious gap in the literature in this sense, too. In addition to this, examining the integration of DevOps practices with all processes, and measuring efficiency, and sharing the results with case studies in which technologies and tools that will deepen this research, which deals with release management as flow, can also enable the expansion of the model and its verification with different researches.